\documentclass[12pt,letterpaper]{article}
\usepackage[includeheadfoot, 
            marginratio={1:1,2:3}, 
            width=412pt, 
            height=688pt,]{geometry}

\usepackage{amsmath}
\usepackage{amsfonts}
\usepackage{amssymb}
\usepackage{bbm}

\usepackage{graphicx}

\allowdisplaybreaks[2]
\numberwithin{equation}{section}

\newcommand{\be}{\begin{equation}}
\newcommand{\ee}{\end{equation}}
\newcommand{\bea}{\begin{eqnarray}}
\newcommand{\eea}{\end{eqnarray}}

\def\IR{\relax{\rm I\kern-.18em R}}

\def\IP{\relax{\rm I\kern-.18em P}}
\def\inbar{\vrule height1.5ex width.4pt depth0pt}
\def\IC{\relax\,\hbox{$\inbar\kern-.3em{\rm C}$}}

\def\K3{{\bf K3}}

\DeclareMathOperator{\sign}{sign}

\begin{document}


\vspace*{-1.5cm}
\begin{flushright}
  {\small
  MPP-2007-163
  }
\end{flushright}

\vspace{1.5cm}
\begin{center}
  {\LARGE
Gauge Thresholds and K\"ahler Metrics for\\[0.2cm]
Rigid Intersecting D-brane Models
  }
\end{center}

\vspace{0.25cm}
\begin{center}
  {\small
  Ralph~Blumenhagen\\
  Maximilian~Schmidt-Sommerfeld\\
  }
\end{center}

\vspace{0.1cm}
\begin{center}
  \emph{
  Max-Planck-Institut f\"ur Physik\\
  F\"ohringer Ring 6\\
  80805 M\"unchen\\
  Germany } \\
\end{center}

\vspace{-0.1cm}
\begin{center}
  \tt{
  blumenha@mppmu.mpg.de\\
  pumuckl@mppmu.mpg.de\\
  }
\end{center}

\vspace{1.5cm}
\begin{abstract}
\noindent
The gauge threshold
corrections for globally consistent $\mathbb{Z}_2\times\mathbb{Z}'_2$ orientifolds 
with rigid intersecting D6-branes are computed. The one-loop corrections
to the holomorphic gauge kinetic function are extracted and the K\"ahler metrics for
the charged chiral multiplets are determined up to two constants.

\end{abstract}

\thispagestyle{empty}
\clearpage
\tableofcontents


\section{Introduction and Motivation}
Gauge threshold corrections in D-brane models
\cite{Uranga:2003pz,MarchesanoBuznego:2003hp,Lust:2004ks,Blumenhagen:2005mu,Blumenhagen:2006ci}
have lately attracted renewed interest. 
This is mainly due to the fact that they were shown to be equal to one-loop amplitudes appearing
in superpotential couplings generated by euclidean D-brane instantons
\cite{Abel:2006yk,Akerblom:2006hx}. 
As the gauge threshold corrections generically are not holomorphic functions of the moduli fields,
it is not a priori clear how these couplings can be incorporated
in a superpotential. It can however be shown that
a cancellation between the non-holomorphic terms in the thresholds and terms arising from
non-trivial, moduli-dependent K\"ahler metrics takes place, thus rendering both the instanton-generated
superpotential and the one-loop corrected gauge kinetic function holomorphic
\cite{Akerblom:2007uc,Billo:2007sw,Billo:2007py}.

Gauge threshold corrections in D6-brane models on toroidal backgrounds have so far
only been computed for so-called bulk branes, which are uncharged under twisted sector RR-fields
\cite{Lust:2003ky,Akerblom:2007np}. 
Here the prototype example is the $\mathbb{Z}_2\times\mathbb{Z}_2$ orbifold
with $h_{21}=3$, which only 
has eight three-cycles, all of which are bulk cycles.
Twisted three-cycles occur for the  $\mathbb{Z}_2\times\mathbb{Z}'_2$ orbifold
with $h_{21}=51$.
This orbifold  is a   particularly interesting
background for intersecting D6-brane models because the CFT is free and thus explicit
calculations can be performed and, most strikingly, there exist rigid
three-cycles. 
These latter ones allow one
to construct models without phenomenologically undesirable adjoint fields\footnote{Similar
constructions without these fields can be made in the Type I string \cite{Dudas:2005jx} or
using shift orientifolds \cite{Antoniadis:1999ux,Blumenhagen:2006ab}.}
and therefore, in particular, asymptotically free gauge theories \cite{Blumenhagen:2005tn}.
Moreover, these rigid cycles are important when studying non-perturbative effects
because euclidean D2-brane (short E2-brane) instantons wrapping 
these cycles have the zero mode structure
needed for  a contribution to the
superpotential \cite{Blumenhagen:2006xt,Cvetic:2007ku}. This means that, due to the
aforementioned relation between one-loop threshold corrections and the one-loop instanton
amplitudes, the results of this paper are relevant when determining E2-instanton effects in toroidal intersecting
D6-brane models.

In this paper, gauge threshold corrections for D6-brane models
on the $\mathbb{Z}_2\times\mathbb{Z}'_2$ orbifold with $h_{21}=51$ \cite{Blumenhagen:2005tn},
in which branes can be charged under the twisted RR fields,
are computed. 
Furthermore, it is shown that also on this background a cancellation between the
non-holomorphic parts of the gauge thresholds and the terms involving the K\"ahler metrics
occurs in an equation relating the holomorphic gauge kinetic function to the physical gauge
coupling, which is the one calculated in string theory.
Since the main body of this paper is quite technical, let us mention
two of our main findings. A summary of all the results including formulas is
given at the end of the paper.
We will determine the one-loop corrections to the holomorphic gauge kinetic
function  $f_a$ for the gauge theory on a brane stack labelled $a$.
It is interesting to see that
on this background, in contradistinction to the case of the
$\mathbb{Z}_2\times\mathbb{Z}_2$ orbifold with $h_{21}=3$
\cite{Akerblom:2007uc}, there are corrections to the gauge
kinetic function from sectors preserving $\mathcal{N}=1$ supersymmetry.

The gauge threshold corrections computed in this paper allow
for a determination of K\"ahler metrics of charged matter on the background considered.
The metric for the vector-like bifundamental matter arising from strings
stretched between two stacks of branes that are coincident but differ in their twisted charges
can be determined using holomorphy arguments and yields the expected results.
Equivalently, one can determine, up to two constants, the K\"ahler metric for
the chiral bifundamental matter arising at the intersection of two stacks of
branes, confirming previous findings.

The results of this paper extend those computed in the T-dual picture \cite{Billo:2007sw,Billo:2007py}
insofar as to be valid in a global rather than local model and to
include more general twisted sector charges. In addition, a new contribution
to the so-called universal gauge coupling corrections
\cite{Derendinger:1991kr,Derendinger:1991hq,Kaplunovsky:1995jw,Akerblom:2007uc} is found.
Their appearance, together with the holomorphy of the gauge kinetic function,
implies a redefinition of the 
twisted complex structure moduli
at one loop.

Before we dwell upon the technical details of our computation, let us spell
out the motivation for this project, which is  twofold. 
Firstly, the gauge threshold corrections are computed. They are important in D-brane models as the gauge couplings
on different branes depend on the volumes of the cycles the branes wrap and therefore are generically
not equal at the compactification or string scale. This poses a potential problem with the apparent
gauge coupling unification seen in the MSSM which could be solved upon taking the gauge threshold corrections
into account.

Secondly, as already mentioned, the present background allows for E2-ins\-tanton contributions to the
superpotential and is thus a good arena to explicitly study non-perturbative effects in intersecting
D6-brane models. Examples of such effects that are important are Majorana mass terms for
the right-handed neutrinos \cite{Cvetic:2007ku,Cvetic:2007qj} and the issue of moduli stabilisation \cite{Camara:2007dy}.
Note in this context that the one-loop corrections determined here lead to a dependence of the instanton-induced terms
on the K\"ahler moduli, whereas, at tree level, they only depend on the complex structure moduli.

\section{Setup and partition functions}
\begin{figure}[ht] 
\begin{center} 
 \includegraphics[width=0.7\textwidth]{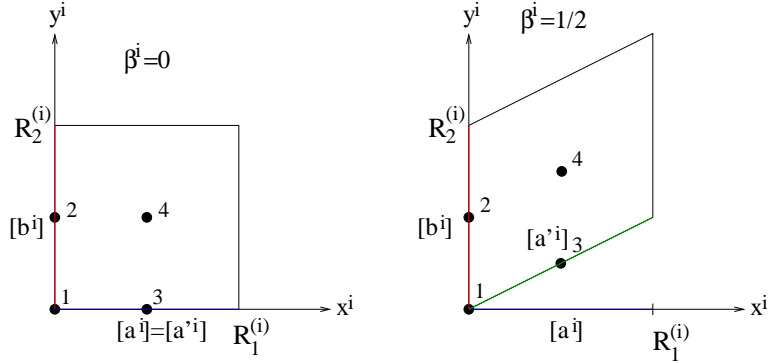} 
\end{center} 
\caption{\small Geometry of the two-tori, orbifold fixed points and one-cycles}\label{figure} 
\end{figure}

The setup considered in this paper \cite{Blumenhagen:2005tn} is an orientifold of an orbifolded torus.
The torus is a factorisable six-torus $T^6=T^2\times T^2 \times T^2$ and the orbifold group is
$\mathbb{Z}_2\times\mathbb{Z}'_2$, where each $\mathbb{Z}_2$-factor inverts two two-tori.
There are three $\mathbb{Z}_2$-twisted sectors with sixteen fixed points each. The orientifold
group is $\Omega R (-1)^{F_L}$, where $\Omega$ is world-sheet parity, $R$ an antiholomorphic involution
and $F_L$ the left-moving spacetime fermion number. The three two-tori have radii $R_{1,2}^{(i)}$
along the $x^i$,$y^i$-axes,
$i\in\{1,2,3\}$. The tori may ($\beta^i=1/2$) or may not ($\beta^i=0$) be tilted. There are four orbifold
fixed points on each torus, at $(0,0)$, $(0,R_2^{(i)}/2)$,  $(R_1^{(i)}/2,\beta^i R_2^{(i)}/2)$ and
$(R_1^{(i)}/2,(1+\beta^i) R_2^{(i)}/2)$. They will be labelled fixed points 1,2,3 and 4. All this geometrical
data is shown in figure \ref{figure} for an untilted and a tilted torus.

(Stacks of) D-branes on this background are described by the wrapping numbers
$(n^i, m^i)$,
the charges under the twisted RR-fields $\epsilon^i\in\{-1,1\}$, the
position $\delta^i\in\{0,1\}$
and the discrete
Wilson lines $\lambda^i\in\{0,1\}$.
The brane wraps the one-cycle $n^i [{a'}^i] + m^i [b^i]$ on the $i$'th torus,
the fundamental one-cycles $[{a'}^i]=[a^i]+\beta^i [b^i]$ and $[b^i]$ are shown in figure \ref{figure}.
The $\epsilon^i$ satisfy $\epsilon^1=\epsilon^2 \epsilon^3$. 
The position is
described by the three parameters $\delta^i$, where $\delta^i=0$ if the brane goes through
fixed point 1 on the i'th torus and $\delta^i=1$ otherwise. An alternative way to characterise a brane is to
use $\epsilon^i_{kl}\in\{-1,0,1\}$, $i\in\{1,2,3\}$, $k,l\in\{1,2,3,4\}$
\cite{Blumenhagen:2005tn}, instead of $\epsilon^i$,
$\delta^i$ and $\lambda^i$. $\epsilon^i_{kl}$ is the charge of the brane under the fixed point labelled $kl$
in the $i$'th twisted sector. The $\epsilon^i_{kl}$ can be determined from $\epsilon^i$,
$\delta^i$ and $\lambda^i$.
Note that for each $i$ only
four out of the sixteen $\epsilon^i_{kl}$ are non-zero.
In both these descriptions there is some redundancy
\cite{Blumenhagen:2005tn}. Rather than fixing some of the $\epsilon^i_{kl}$ charges
to be 1 \cite{Blumenhagen:2005tn}, it will here be more convenient to
choose $n^{1,2} > 0$ (or $m^i$ positive if $n^i$ vanishes).

It is useful
to define $\widetilde{m}^i = m^i + \beta^i n^i$, such that a brane wraps
the one-cycle $n^i [a^i] + \widetilde{m}^i [b^i]$ on the $i$'th torus.
The volume of this one-cycle
is given by
\begin{eqnarray}
 V^i=\sqrt{(n^i)^2 (R_1^{(i)})^2 + (\widetilde{m}^i)^2 (R_2^{(i)})^2}
 \label{onecyclevolume}
\end{eqnarray}
and the tree-level gauge coupling reads:
\begin{eqnarray}
 \frac{1}{g_{tree}^2} = e^{-\phi_{10}} \prod_i V^i
 = \frac{e^{-\phi_{4}}}{(T_1T_2T_3)^{1/2}} \prod_i V^i
 = \frac{(SU_1U_2U_3)^{1/4}}{(T_1T_2T_3)^{1/2}} \prod_i V^i\, .
 \label{treelevelgc}
\end{eqnarray}
Here, $\phi_{10}$($\phi_{4}$) is the 10(4)-dimensional dilaton in string frame,
$S$ is the dilaton in Einstein frame, $U_i$ are the (real parts of the) complex structure moduli in Einstein frame and
$T_i = R_1^{(i)} R_2^{(i)}$ are the (real parts of the) K\"ahler moduli. From \eqref{treelevelgc}
one can, using the supersymmetry condition (see below), derive the dependence of the
tree-level gauge kinetic function on the untwisted moduli \cite{Lust:2004cx}:
\begin{eqnarray}
 \hat{f}_{tree} = S^c n^1 n^2 n^3 - \sum_{i\neq j \neq k=1}^3 U_i^c n^i
 \widetilde{m}^j \widetilde{m}^k \, .
 \label{treelevelgkf}
\end{eqnarray}
$S^c$ and $U_i^c$ are the complexified dilaton and complex structure moduli, the axions being RR-fields.
Similarly, $T^c_i$ are complexified K\"ahler moduli, the axions stemming from the NSNS 2-form-field.

The D-brane is rotated by the angles $\theta^i$, defined via $\tan\theta^i=\widetilde{m}^i R_2^{(i)}/n^i R_1^{(i)}$, with respect to the x-axes of the three tori. Only supersymmetric
configurations will be considered in this paper, i.e. $\sum_{i=1}^3 \theta^i =0$.\footnote{Branes
at angles $\theta^i$ satisfying $\sum_{i=1}^3 \theta^i =\pm 2\pi$ are also supersymmetric, but are,
for simplicity, not considered here.}

The charges of the four orientifold planes are denoted $\eta_{\Omega R}$ and $\eta_{\Omega R i}$,
$i\in \{1,2,3\}$ and have to satisfy
\begin{eqnarray}
 \eta_{\Omega R} \prod_{i=1}^3 \eta_{\Omega R i} = -1
\end{eqnarray}
in the present case of the $\mathbb{Z}_2\times\mathbb{Z}'_2$ orbifold with $h_{21}=51$ \cite{Blumenhagen:2005tn}.
The tadpole cancellation conditions are given by
\begin{eqnarray}
 && \sum_a N_a n_a^1 n_a^2 n_a^3 = 16 \eta_{\Omega R} \\ &&
 \sum_a N_a n_a^i \widetilde{m}_a^j \widetilde{m}_a^k = -2^{4-2\beta^j-2\beta^k} \eta_{\Omega R i}
 \qquad i\neq j\neq k \in \{1,2,3\} \\ &&
 \sum_a N_a n_a^i (\epsilon_{a,kl}^i -\eta_{\Omega R} \eta_{\Omega R i} \epsilon_{a,R(k)R(l)}^i) = 0
 \label{twtadp1} \\ &&
 \sum_a N_a \widetilde{m}_a^i (\epsilon_{a,kl}^i + \eta_{\Omega R} \eta_{\Omega R i} 
      \epsilon_{a,R(k)R(l)}^i) = 0 \label{twtadp2} ,
\end{eqnarray}
where $R(k)=k$ in case of an untilted torus and $R(\{ 1,2,3,4\})=\{1,2,4,3\}$ in the other case
\cite{Blumenhagen:2005tn} and the sum is a sum over all stacks of branes. $N_a$ denotes
the number of branes on stack $a$. The wrapping numbers and twisted charges carry an
index $a$ denoting the brane stack which they describe.
The orientifold projection acts on the wrapping numbers and twisted charges as follows:
\begin{eqnarray}
 \widetilde{m}^I &\rightarrow& - \widetilde{m}^I \\
 \epsilon_{kl}^i &\rightarrow& - \eta_{\Omega R} \eta_{\Omega R i}
 \epsilon^i_{R(k)R(l)}\, .
\end{eqnarray}

The massless open string spectrum can be read of from the open string partition function
given by annulus (and M\"obius) amplitudes without any vertex operators inserted.
These can be calculated from boundary (and crosscap) states
describing the D-branes (and O-planes)
\cite{Gaberdiel:1999ch,Gaberdiel:2000jr,Stefanski:2000fp,Gaberdiel:2000fe,Quiroz:2001xz,Craps:2001xw,Maiden:2006qe}.
Only the annulus amplitudes will be given here, and four cases will be distinguished
(These are not all possibilities there are, but all needed here.), that will also be
important in the rest of this paper:

\paragraph{Case 1:} The annulus has both boundaries on the same stack of
branes $a$. The amplitude is
\begin{eqnarray}
 \mathcal{A}_{aa}^{(1)}  &=& - N_a^2 \int_0^\infty dl \Bigg[
      \frac{\vartheta_3^4-\vartheta_4^4-\vartheta_2^4+\vartheta_1^4}{\eta^{12}}
      \prod_{i=1}^3 \frac{(V_a^i)^2}
               {R_1^{(i)} R_2^{(i)}} \widetilde{L}_{aa}^{(i)} \nonumber \\ &&
      + 16 \sum_{i=1}^3 \sigma_{aa}^i
      \frac{\vartheta_3^2 \vartheta_2^2 - \vartheta_4^2 \vartheta_1^2 -
                     \vartheta_2^2 \vartheta_3^2 + \vartheta_1^2 \vartheta_4^2}{\eta^6 \vartheta_4^2}
      \frac{(V_a^i)^2}
               {R_1^{(i)} R_2^{(i)}} \widetilde{L}_{aa}^{(i)} \Bigg],
\end{eqnarray}
where $\sigma_{ab}^i = \frac{1}{4} \sum_{k,l=1}^4 \epsilon^i_{a,kl} \epsilon^i_{b,kl}$,
$\vartheta=\vartheta(0,2il)$, $\eta=\eta(2il)$ and
$V_a^i$, defined in \eqref{onecyclevolume}, now carries an index $a$ to denote the brane stack considered. The
following Kaluza-Klein and winding sum has been defined \cite{Blumenhagen:1999md}:
\begin{eqnarray}
 \widetilde{L}_{ab}^{(i)} &=& \sum_{m,w} \exp \left[ -\pi l
       (V_a^i)^2
       \left( \frac{m^2}{(R_1^{(i)} R_2^{(i)})^2} + w^2 \right) 
       + i \pi m (\delta_a^i-\delta_b^i) + i \pi w (\lambda_a^i-\lambda_b^i)
       \right] \nonumber
\end{eqnarray}

\paragraph{Case 2:} The annulus stretches between two stacks of D-branes $a$ and $b$
wrapping the same submanifold of the covering six-torus of the internal space and having the same
discrete Wilson lines turned on($\lambda_a^i=\lambda_b^i$). This means in particular
$\theta_a^i=\theta_b^i$,$V_a^i=V_b^i,\delta_a^i=\delta_b^i$, however not all
twisted charges $\epsilon^i$ are equal. 

One can show that in this case
$\sigma_{ab}^i =\pm 1$.
The amplitude is
\begin{eqnarray}
 \mathcal{A}_{ab}^{(2)}  &=& - N_a N_b \int_0^\infty dl \Bigg[
      \frac{\vartheta_3^4-\vartheta_4^4-\vartheta_2^4+\vartheta_1^4}{\eta^{12}}
      \prod_{i=1}^3 \frac{(V_a^i)^2}
               {R_1^{(i)} R_2^{(i)}} \widetilde{L}_{ab}^{(i)} \nonumber \\ &&
      + 16 \sum_{i=1}^3 \sigma_{ab}^i
      \frac{\vartheta_3^2 \vartheta_2^2 - \vartheta_4^2 \vartheta_1^2 -
                     \vartheta_2^2 \vartheta_3^2 + \vartheta_1^2 \vartheta_4^2}{\eta^6 \vartheta_4^2}
      \frac{(V_a^i)^2}
               {R_1^{(i)} R_2^{(i)}} \widetilde{L}_{ab}^{(i)} \Bigg].
\end{eqnarray}
\paragraph{Case 3:} The two stacks of branes wrap submanifolds that are homologically equal
on the covering torus (This implies $\theta_a^i=\theta_b^i$,$V_a^i=V_b^i$.),
but do not satisfy $\lambda_a^i=\lambda_b^i$, $\delta_a^i=\delta_b^i$ for all $i$. The amplitude looks
as the one of case 2.
\paragraph{Case 4:} The annulus stretches between two branes that
intersect at non-trivial angles on all three tori. The amplitude reads
\begin{eqnarray}
 && \mathcal{A}_{ab}^{(4)}  = N_a N_b \int_0^\infty dl 
         \Bigg[ 8 \left({\textstyle \prod_{i=1}^3 I_{ab}^i}\right)
         \sum_{\alpha,\beta} (-1)^{2(\alpha+\beta)} \frac{\vartheta\genfrac[]{0pt}{}{\alpha}{\beta}(0)}{\eta^3}
         \prod_{i=1}^3 \frac{\vartheta\genfrac[]{0pt}{}{\alpha}{\beta} (\theta_{ab}^i)     }
               {\vartheta\genfrac[]{0pt}{}{1/2}{1/2} (\theta_{ab}^i)}   \\ &&
         + 32 \sum_{i\ne j\ne k}  I_{ab}^i\, \sigma_{ab}^i
         \sum_{\alpha,\beta} (-1)^{2(\alpha+\beta)} 
         \frac{\vartheta\genfrac[]{0pt}{}{\alpha}{\beta}(0)}{\eta^3}
         \frac{\vartheta\genfrac[]{0pt}{}{\alpha}{\beta}(\theta_{ab}^i)}{\vartheta\genfrac[]{0pt}{}{1/2}{1/2}(\theta_{ab}^i)}
         \frac{\vartheta\genfrac[]{0pt}{}{|\alpha-1/2|}{\beta}(\theta_{ab}^j)}{\vartheta\genfrac[]{0pt}{}{0}{1/2}(\theta_{ab}^j)}
         \frac{\vartheta\genfrac[]{0pt}{}{|\alpha-1/2|}{\beta}(\theta_{ab}^k)}{\vartheta\genfrac[]{0pt}{}{0}{1/2}(\theta_{ab}^k)}
         \Bigg], \nonumber 
\end{eqnarray}
where 
$\theta_{ab}^i=\theta_a^i-\theta_b^i$ and the intersection number $I_{ab}^i=(\widetilde{m}_a^{i} n_b^{i} - \widetilde{m}_b^{i} n_a^{i})$
were defined.

Upon transforming into the open string channel, one finds the following massless spectrum:
Case 1 yields a massless vector multiplet in the adjoint representation of $U(N_a)$.
Case 2 gives a massless hypermultiplet in the bifundamental representation of
$U(N_a)\times U(N_b)$. In case 3 there are no massless fields. In case 4 one finds $|\Upsilon_{ab}|$,
$\Upsilon_{ab}=\frac{1}{4}\prod_{i=1}^3 I_{ab}^i + \sum_{i=1}^3 I_{ab}^i \sigma_{ab}^i$,
chiral multiplets in the bifundamental representation of
$U(N_a)\times U(N_b)$.

\section{Gauge threshold corrections}

The gauge threshold corrections will be computed using the background field
method employed previously \cite{Bachas:1996zt,Antoniadis:1999ge,Lust:2003ky}. This means that
the one-loop correction to the gauge coupling of brane $a$ induced by brane $b$ is determined as
follows. First, one replaces the 4d spacetime part of the partition function in the above
expression as follows \cite{Lust:2003ky}:
\begin{eqnarray}
 \frac{\vartheta\genfrac[]{0pt}{}{\alpha}{\beta}(0,2il)}{\eta^3(2il)}
 \rightarrow 2 i \pi B q_a
  \frac{\vartheta\genfrac[]{0pt}{}{\alpha}{\beta}(-\epsilon_a,2il)}
       {\vartheta\genfrac[]{0pt}{}{1/2}{1/2}(-\epsilon_a,2il)}, 
\end{eqnarray}
where $\pi \epsilon_a=\arctan(\pi q_a B)$, $q_a$ is the charge of the open string ending on brane $a$
and $B$ is the background magnetic field in the 4d spacetime. One then expands the resulting
expressions in a series in $B$ and the coefficient of $B^2$ gives the desired
expression for the correction to the gauge coupling, which needs to be evaluated further.

For the four cases distinguished above one finds, denoting the amplitudes after the manipulations described with an
additional superscript $g$:
\begin{eqnarray}
 \mathcal{A}_{aa}^{g(1)} &=& 32 \pi\, N_a^2  \int_0^\infty dl 
      \sum_{i=1}^3 \sigma_{aa}^i
      \frac{(V_a^i)^2}
               {R_1^{(i)} R_2^{(i)}} \widetilde{L}_{aa}^{(i)}
 \label{gccorr1} \\
 \mathcal{A}_{ab}^{g(2)} &=& 32 \pi\, N_a N_b  \int_0^\infty dl 
      \sum_{i=1}^3 \sigma_{ab}^i
      \frac{(V_a^i)^2}
               {R_1^{(i)} R_2^{(i)}} \widetilde{L}_{ab}^{(i)} 
 \label{gccorr2} \\
\mathcal{A}_{ab}^{g(3)} &=& 32 \pi\, N_a N_b  \int_0^\infty dl 
      \sum_{i=1}^3 \sigma_{ab}^i
      \frac{(V_a^i)^2}
               {R_1^{(i)} R_2^{(i)}} \widetilde{L}_{ab}^{(i)} 
 \label{gccorr3} \\
 \mathcal{A}_{ab}^{g(4)} &=& N_a N_b 
 \int_0^\infty dl \Bigg[ 8 \left({\textstyle \prod_{i=1}^3 I_{ab}^i}\right)\, \sum_{i=1}^3
 \frac{\vartheta'_1 (\theta_{ab}^i,2il)}{\vartheta_1(\theta_{ab}^i,2il)}
 + \sum_{i\ne j\ne k} 32 I_{ab}^i\,  \sigma_{ab}^i \nonumber \\ &&
 \left( \frac{\vartheta'_1(\theta_{ab}^i,2il)}{\vartheta_1(\theta_{ab}^i,2il)} 
 +  \frac{\vartheta'_4(\theta_{ab}^j,2il)}{\vartheta_4(\theta_{ab}^j,2il)}+
    \frac{\vartheta'_4(\theta_{ab}^k,2il)}{\vartheta_4(\theta_{ab}^k,2il)} \right)
 \Bigg] .\label{gccorr4}
\end{eqnarray}
The overall normalisation will later be fixed by demanding the  running of the gauge
coupling with the correct beta function coefficient, but the relative
normalisation is taken into account correctly. The full correction to the gauge coupling
on brane $a$ is given by summing over all annuli (and M\"obii) with at least one
boundary on brane $a$.

The above expressions can be evaluated analogously to  the corresponding
ones in the $\mathbb{Z}_2\times\mathbb{Z}_2$  orbifold with $h_{21}=3$
\cite{Lust:2003ky,Akerblom:2007np}.
The results are:
\begin{eqnarray}
 \mathcal{A}_{aa}^{g(1)} &=& 32 \pi\, N_a^2 \int_0^\infty dl 
      \sum_{i=1}^3 \sigma_{aa}^i
      \frac{(V_a^i)^2}{R_1^{(i)} R_2^{(i)}}
  \label{twtadpole1} \\ &&
      + 32 \pi\, N_a^2 \left( \sigma_{aa} \ln\left[ \frac{M_s^2}{\mu^2}\right] 
      - \sum_{i=1}^3 \sigma_{aa}^i \ln\left[ (V_a^i)^2\right] \right)
  \label{annurunning1annukaehlermet1} \\ &&
      - 32 \pi\, N_a^2 \sum_{i=1}^3 4\, \sigma_{aa}^i \, \ln\left[ \eta(iR_1^{(i)} R_2^{(i)})\right] 
  \label{holocorr1} \\ &&
      - 32 \pi\,  N_a^2 \sigma_{aa} \ln [4\pi]\, .
  \label{constants1}
\end{eqnarray}
Here, $\sigma_{ab}=\sum_{i=1}^3 \sigma_{ab}^i$ was defined and
the divergence $\int^\infty \frac{dt}{t}$ arising from the massless open string modes was
replaced by $\ln\left[ M_s^2/\mu^2\right]$ \cite{Akerblom:2007np}.
For the cases 2 and 3 we obtain
\begin{eqnarray}
 \mathcal{A}_{ab}^{g(2)} &=& 32 \pi\,  N_a N_b \int_0^\infty dl 
      \sum_{i=1}^3 \sigma_{ab}^i
      \frac{(V_a^i)^2}{R_1^{(i)} R_2^{(i)}}
  \label{twtadpole2} \\ &&
      + 32 \pi\,  N_a N_b
      \sum_{i=1}^3 \sigma_{ab}^i \left( \ln\left[ \frac{M_s^2}{\mu^2}\right]
      - \ln\left[ (V_a^i)^2\right]
      - 4 \ln\left[ \eta(iR_1^{(i)} R_2^{(i)})\right] \right) \nonumber \\
  \label{annurunning2annukaehlermet2holocorr2} \\ &&
      - 32 \pi N_a N_b \sigma_{ab} \ln [4\pi]
  \label{constants2} \\
      \mathcal{A}_{ab}^{g(3)} &=& 32 \pi N_a N_b \int_0^\infty dl 
      \sum_{i=1}^3 \sigma_{ab}^i
      \frac{(V_a^i)^2}{R_1^{(i)} R_2^{(i)}}
  \label{twtadpole3} \\ &&
      - 64 \pi N_a N_b \sum_{i=1}^3 \sigma_{ab}^i
      \ln\left[ \frac{\vartheta
      \genfrac[]{0pt}{}{1/2(1-|\delta_a^i-\delta_b^i|)}{1/2(1-|\lambda_a^i-\lambda_b^i|)}
      (0,iR_1^{(i)} R_2^{(i)})}{\eta(iR_1^{(i)} R_2^{(i)})}\right] \, .
  \label{holocorr3} 
\end{eqnarray}
Note that in case 2 there is again a divergence proportional to $\int^\infty \frac{dt}{t}$,
which was replaced by $\ln [M_s^2/\mu^2]$, whereas
there is none in case 3 due to the absence of massless modes in this sector.
Finally, for case 4 the thresholds are
\begin{eqnarray}
 \mathcal{A}_{ab}^{g(4)} &=& 
        N_a N_b \int_0^\infty dl \ 8 \left({\textstyle \prod_{i=1}^3 I_{ab}^i}\right) \sum_{i=1}^3 \pi 
 \cot\left[\pi \theta_{ab}^i\right]
    \label{untwtadpole} \\ &&
        + N_a N_b \int_0^\infty dl \sum_{i=1}^3 32\,  I_{ab}^i\,
        \sigma_{ab}^i\, 
        \pi \cot\left[\pi\theta_{ab}^i\right]
    \label{twtadpole4} \\ &&
        + 16 \pi\, N_a N_b \Upsilon_{ab} \sum_{i=1}^3 \left( s_{ab}^i \ln\left[ \frac{M_s^2}{\mu^2}\right] +
              \ln \left[ \frac{\Gamma(1-|\theta_{ab}^i|)}
              {\Gamma(|\theta_{ab}^i|)}\right]^{s_{ab}^i} \right)
    \label{annurunning4annukaehlermet4} \\ &&
        + 64 \pi\,  N_a N_b \ln [2] \sum_i I_{ab}^i\,
        (\theta_a^i-\theta_b^i)\,  \sigma_{ab}^i
    \label{annuuni4} \\ &&
        + 16 \pi N_a N_b \left( (\ln [2] - \gamma_E) \Upsilon_{ab} \sum_i s_{ab}^i
        + \ln [4] \sum_{i\ne j\ne k} I_{ab}^i \sigma_{ab}^i ( s_{ab}^j + s_{ab}^k )
        \right)\, , \nonumber \\
    \label{constants4}
\end{eqnarray}
where the abbreviation $s_{ab}^i=\sign(\theta_{ab}^i)$ was used.

The contributions \eqref{constants1}, \eqref{constants2} and \eqref{constants4}
are just moduli independent,
finite constants and as such of no further interest.
The terms in \eqref{twtadpole1}, \eqref{twtadpole2}, \eqref{twtadpole3}, \eqref{untwtadpole} and
\eqref{twtadpole4} are divergent integrals and the sum over these contributions from all annuli
has to cancel. The expression \eqref{untwtadpole} also appears in the model
on the orbifold with $h_{21}=3$ and the sum can be shown to vanish upon using the
untwisted tadpole cancellation conditions \cite{Lust:2003ky}
\footnote{For this to happen, the M\"obius amplitudes have to be taken into account as well,
but they are unchanged as the orientifold planes do not carry twisted charges.}.

Using
$(V_a^i)^2=(n_a^i)^2 (R_1^{(i)})^2 + (\widetilde{m}_a^i)^2 (R_2^{(i)})^2$, 
$\tan\theta_a^i=\widetilde{m}_a^i R_2^{(i)}/n_a^i R_1^{(i)}$ as well as the
fact that in cases 1 and 2 $(n_a^i,\widetilde{m}_a^i)=(n_b^i,\widetilde{m}_b^i)$
the three terms \eqref{twtadpole1}, \eqref{twtadpole2}, \eqref{twtadpole3} and
\eqref{twtadpole4} can all be brought into the form\footnote{Note
that due to the above choice of signs for $n^{1,2}$, $m^{1,2}$ and
the supersymmetry condition, not only the absolute values but also the signs
of the wrapping numbers are equal.}
\begin{eqnarray}
\mathcal{A}_{ab}^{TT} =
 8 \pi N_a N_b \int_0^\infty dl \sum_{i=1}^3 \sum_{k,l=1}^4 \epsilon^i_{a,kl} \epsilon^i_{b,kl}
      \frac{n_a^i n_b^i (R_1^{(i)})^2 + \widetilde{m}_a^i \widetilde{m}_b^i
 (R_2^{(i)})^2}{R_1^{(i)} R_2^{(i)}}\, ,
\end{eqnarray}
where the superscript $TT$ denotes that these are the contributions from \eqref{gccorr1}, \eqref{gccorr2},
\eqref{gccorr3} and \eqref{gccorr4} that vanish after using the
twisted tadpole cancellation conditions.
Summing over this contribution from all annuli yields (denoting the orientifold
image of brane $b$ by $b'$)
\begin{eqnarray}
 && \sum_{b\neq a} (\mathcal{A}_{ab}^{TT} + \mathcal{A}_{ab'}^{TT}) + \mathcal{A}_{aa'}^{TT} \\
 &=& N_a \sum_i \frac{n_a^i}{R_1^{(i)} R_2^{(i)}} \sum_{k,l} \epsilon^i_{a,kl}
 \Bigg[ (R_1^{(i)})^2 \sum_b N_b n_b^i ( \epsilon^i_{b,kl} - \eta_{\Omega R} \eta_{\Omega R i}
 \epsilon^i_{b,R(k)R(l)} ) \Bigg] \nonumber \\ && 
 + N_a \sum_i \frac{\widetilde{m}_a^i}{R_1^{(i)} R_2^{(i)}} \sum_{k,l} \epsilon^i_{a,kl}
 \Bigg[ (R_2^{(i)})^2 \sum_b N_b \widetilde{m}_b^i ( \epsilon^i_{b,kl} + \eta_{\Omega R} \eta_{\Omega R i}
 \epsilon^i_{b,R(k)R(l)} ) \Bigg] \nonumber ,
\end{eqnarray}
which vanishes upon using the twisted tadpole conditions \eqref{twtadp1} and \eqref{twtadp2}.
The other contributions to the gauge coupling corrections will be discussed in
the next sections.
\section{Holomorphic gauge kinetic function}
In a  supersymmetric gauge theory one can compute the running 
gauge couplings $g_a(\mu^2)$ in terms of the gauge kinetic functions $f_a$, the 
K\"ahler potential ${\cal K}$ and the K\"ahler metrics of the charged matter 
fields $K^{ab}(\mu^2)$ 
\cite{Shifman:1986zi,Kaplunovsky:1994fg,Kaplunovsky:1995jw}: 
\bea 
\label{gceffFT} 
 \frac{8\pi^2}{g_a^2(\mu^2)} &=& 8\pi^2\,\Re(f_a)
 + \Delta_0 + \sum_r \Delta_r
\eea 
with
\begin{eqnarray}
 \Delta_0 &=& T(G_a) \left( - \frac{3}{2} \ln\left[ \frac{\Lambda^2}{\mu^2}\right] 
                         - \frac{1}{2} {\cal K} + \ln\left[ \frac{1}{g_{a}^2(\mu^2)}\right] \right)
 \\
 \Delta_r &=& T_a(r) \left( \frac{n_r}{2} \ln\left[
 \frac{\Lambda^2}{\mu^2}\right] + 
\frac{n_r}{2} {\cal K} - \ln\left[ \det K_r (\mu^2)\right] \right)
\end{eqnarray}
and $T_a(r) = {\rm Tr} (T^2_{(a)})$ ($T_{(a)}$ being the 
generators of the gauge group $G_a$). 
In addition, $T(G_a) = T_a({\rm adj}\, G_a)$ and $n_r$ is the 
number of multiplets in the representation $r$ of the gauge group and the  
sum in \eqref{gceffFT} runs over these representations.
For this paper only the gauge groups $SU(N_a)$ and its fundamental, adjoint, symmetric and antisymmetric
representations ($r=f,adj,s,a$) are relevant. For these $T(G_a) = T_a(adj)=N_a$, $T_a(f)=1/2$, $T_a(s)=(N_a+2)/2$ and $T_a(a)=(N_a-2)/2$.
In this context, the natural cutoff scale for a field theory is the Planck scale, 
i.e. $\Lambda^2=M_{\mathrm{Pl}}^2$.  

The stringy one-loop correction to the LHS of \eqref{gceffFT} was calculated in
the previous section. As, in a supersymmetric theory,
$f_a$ is a holomorphic function of the chiral
superfields, the non-holomorphic terms on the LHS of \eqref{gceffFT} better
be equal to the non-holomorphic terms in $\Delta_0$ and $\Delta_r$.
It will be shown that
this is actually the case for the model under consideration, apart from some universal threshold corrections
\cite{Derendinger:1991kr,Derendinger:1991hq,Kaplunovsky:1995jw,Akerblom:2007uc}
to be discussed in the next section.

$\Delta_0$ must match
the contribution of the annulus $\mathcal{A}_{aa}^{g(1)}$ to the LHS of \eqref{gceffFT}. On the orbifold
with $h_{21}=3$, the latter vanishes \cite{Lust:2003ky} and the terms
on the RHS cancel amongst each other \cite{Akerblom:2007uc}. In the present case,
things are a little bit different. There are no chiral multiplets in the adjoint
of the gauge group such that,
using ${\cal K}=-\ln(SU_1U_2U_3T_1T_2T_3)$, $M_{\mathrm{Pl}}^2\varpropto M_s^2\sqrt{SU_1U_2U_3}$ and
\eqref{treelevelgc}, 
the one loop contribution in $\Delta_0$
becomes
\begin{eqnarray}
 \frac{\Delta_0}{T(G_a)} =
 -\frac{3}{2} \ln\left[ \frac{M_{\mathrm{Pl}}^2}{\mu^2} \right]- \frac{1}{2} {\cal K} + \ln\left[ \frac{1}{g_{a, tree}^2}\right]
 = -\frac{3}{2} \ln\left[ \frac{M_s^2}{\mu^2}\right] + \ln\left[ \prod_i
 V^i_a\right]\, .
 \label{adjcontrRHS}
\end{eqnarray}

The RHS of \eqref{adjcontrRHS}  matches (up to the overall normalisation, the relative
normalisation of the two terms is however correct) precisely the terms
\eqref{annurunning1annukaehlermet1}. The term \eqref{holocorr1} has
no corresponding one on the RHS of \eqref{gceffFT}, but upon
complexifying the K\"ahler moduli $T_i\rightarrow T^c_i$ (the axions stem
from the NSNS 2-form-field) it can be analytically continued
to a holomorphic function of the complex K\"ahler moduli. One is thus
lead to conclude that there is a one-loop correction to
the gauge kinetic function of the form
\begin{eqnarray}
 \delta_a f_a^{1-loop} =
 \frac{N_a}{4\pi^2} \sum_{i=1}^3 \ln\left[ \eta(i T_i^c)\right]\,  ,
 \label{gkf1loop1}
\end{eqnarray}
especially as there is a very similar correction
in the case of the $\mathbb{Z}_2\times\mathbb{Z}_2$ orbifold with $h_{21}=3$
arising there from a different
open string sector \cite{Akerblom:2007uc}.
The normalisation of the term on the RHS of \eqref{gkf1loop1} is determined from the relative normalisation
of the different terms in \eqref{gceffFT} and \eqref{twtadpole1}, \eqref{annurunning1annukaehlermet1}, \eqref{holocorr1}, \eqref{constants1}.

Next, the contributions in case 2 will be discussed.
Note that in this case $|\sigma_{ab}^i|=|\sigma_{ab}|=1$.
The terms in \eqref{annurunning2annukaehlermet2holocorr2} give a contribution to
the LHS of \eqref{gceffFT} proportional to
\begin{eqnarray}
 \mathcal{A'}_{ab}^{g(2)} &=&
 32 \pi N_a N_b \sigma_{ab} \Bigg( \ln\left[ \frac{M_s^2}{\mu^2}\right]
 \nonumber \\ &&
 - 2 \sigma_{ab} \ln\left[ \prod_{i=1}^3 (V^i_a)^{\sigma_{ab}^i}\right]
 - 4 \sigma_{ab} \sum_{i=1}^3 \sigma_{ab}^i \ln \left[\eta(iT_i)\right] \Bigg).
 \label{case2raw}
\end{eqnarray}
where the prime on $\mathcal{A'}_{ab}^{g(2)}$ denotes that the tadpole and constant contributions
have been subtracted.

This is to be compared (up to the overall normalisation) with the contribution of $n_f=2 N_b$ multiplets
in the fundamental representation ($T_a(f)=1/2$) of the gauge group
to the RHS of \eqref{gceffFT}:
\begin{eqnarray}
 \Delta_f =
 \frac{2 N_b}{4} \left( \ln\left[ \frac{M_s^2}{\mu^2}\right] - 2 \ln \left[ K_f (SU_1U_2U_3)^\frac{1}{4}
 (T_1T_2T_3)^\frac{1}{2} \right] \right)\, .
\end{eqnarray}
One concludes that the K\"ahler metric for the two chiral multiplets in this sector is
\begin{eqnarray}
 K_f^V = (SU_1U_2U_3)^{-\frac{1}{4}} (T_1T_2T_3)^{-\frac{1}{2}} 
       \left( \prod_{i=1}^3 (V^i_a)^{\sigma_{ab}^i} \right)^\frac{1}{\sigma_{ab}} ,
 \label{kaehlermetrichyper}
\end{eqnarray}

and that there is the following one-loop correction to the gauge kinetic function:
\begin{eqnarray}
 \delta_{b^{(2)}} f_a^{1-loop} =
 -\frac{1}{4\pi^2} \sum_b N_b\, \sigma_{ab}\,
                   \sum_{i=1}^3 \sigma^i_{ab} \ln\left[ \eta(iT^c_i)\right]\, .
 \label{gkf1loop2}
\end{eqnarray}
The normalisation of the terms on the RHS of \eqref{gkf1loop2} is determined from the relative normalisation of the
different terms in \eqref{gceffFT} and \eqref{case2raw}.
There is an overall minus sign in \eqref{gkf1loop2} as compared to \eqref{gkf1loop1} which essentially comes from
the fact that the gauge multiplet itself contributes with a different sign to the beta function than chiral multiplets.
Upon changing variables, the K\"ahler metric \eqref{kaehlermetrichyper}
is identical to the one for adjoint fields in the model with $h_{21}=3$
\cite{Lust:2004cx,Blumenhagen:2006ci} as one would expect from the fact that these
fields are described by the same vertex operators in the worldsheet CFT.

The contribution from case 3 to the LHS of \eqref{gceffFT} is
finite after using the tadpole cancellation condition as one would expect from the fact
that there are no massless open string states in this sector. One concludes that
the term \eqref{holocorr3} leads to the following correction to the
gauge kinetic function:
\begin{eqnarray}
 \delta_{b^{(3)}} f_a^{1-loop} =
 -\frac{1}{8\pi^2} \sum_b \frac{N_b}{\sigma_{ab}}
                   \sum_{i=1}^3 \sigma^i_{ab}\, \ln\left[
      \frac{\vartheta
      \genfrac[]{0pt}{}{1/2(1-|\delta_a^i-\delta_b^i|)}{1/2(1-|\lambda_a^i-\lambda_b^i|)}
      (0,iT^c_i)}{\eta(iT^c_i)}\right] 
\end{eqnarray}
Finally, there is the sector yielding the chiral bifundamentals (case 4).
The terms in \eqref{annurunning4annukaehlermet4} contribute
\begin{eqnarray}
 \mathcal{A'}_{ab}^{g(4)} &=& 
 16 \pi N_a N_b \Upsilon_{ab} \sum_i \sign(\theta_{ab}^i) 
 \Bigg( \ln\left[ \frac{M_s^2}{\mu^2}\right] \nonumber \\ && + 
 \frac{1}{\sum_j \sign(\theta_{ab}^j)}
        \ln\left[ \prod_{k=1}^3 \left( \frac{\Gamma(1-|\theta_{ab}^k|)}
              {\Gamma(|\theta_{ab}^k|)}\right)^{\sign(\theta_{ab}^k)}\right] \Bigg)\, ,
\end{eqnarray}
where the prime in $\mathcal{A'}_{ab}^{g(4)}$ denotes omission of the tadpole and constant contributions
and those from \eqref{annuuni4},
to the LHS of \eqref{gceffFT}. An equal contribution
(up to the overall normalisation) to the
RHS results, if the K\"ahler metric for the chiral bifundamentals
is:
\begin{eqnarray}
 K_{f,ab}^{C(1)}
 &=& (SU_1U_2U_3)^{-\frac{1}{4}} (T_1T_2T_3)^{-\frac{1}{2}} \, 
   \left[ \prod_{i=1}^3 \left( \frac{\Gamma(1-|\theta_{ab}^i|)}
              {\Gamma(|\theta_{ab}^i|)}\right)^{\sign(\theta_{ab}^i)}
 \right]^{-1/[2\sum_j \sign(\theta_{ab}^j)]}  \, .
 \label{kaehlermetbifun1}
\end{eqnarray}
Note that this agrees with the result obtained in the case with $h_{21}=3$
\cite{Akerblom:2007uc}, it is however more general in that it allows for arbitrary
signs of $\theta_{ab}^j$.

\section{Universal gauge coupling corrections}

At first sight it might seem that holomorphy implies that the exact K\"ahler
metric for the chiral bifundamentals is given by \eqref{kaehlermetbifun1}.
However, this is not true. As in the case of the
$\mathbb{Z}_2\times\mathbb{Z}_2$ orbifold with $h_{21}=3$
\cite{Akerblom:2007uc}, an additional factor is allowed if, at the same time, the
dilaton and complex structure moduli are redefined at one-loop. This redefinition is
related to sigma-model anomalies in the low energy supergravity theory
\cite{Derendinger:1991kr,Derendinger:1991hq}.

The factor takes the form \cite{Blumenhagen:2006ci,Lust:2004cx,Akerblom:2007uc}
\begin{eqnarray}
 K_{f,ab}^{C(2)} =
 \prod_{i=1}^3 U_i^{-\xi \sign(\Upsilon_{ab}) \theta_{ab}^i}
               T_i^{-\zeta \sign(\Upsilon_{ab}) \theta_{ab}^i}
 \label{kaehlermetfac2},
\end{eqnarray}
where $\xi$ and $\zeta$ are undetermined constants. Upon summing over all chiral
matter charged under $U(N_a)$ and using both the twisted and untwisted tadpole cancellation conditions,
this term leads to the following contribution
to the RHS of \eqref{gceffFT} \cite{Akerblom:2007uc}:
\begin{eqnarray}
 && - {\sum_b}' T_a(f) \left( \ln \det K_{f,ab}^{C(2)} + \ln \det K_{f,ab'}^{C(2)} \right)
 \nonumber \\ &&
 - T_a(a) \ln \det K_{f,aa'}^{C(2)}
 - T_a(s) \ln \det K_{f,aa'}^{C(2)}
 \nonumber \\
 &=& - \frac{1}{4} n_a^1 n_a^2 n_a^3 \left[ \sum_b N_b \widetilde{m}_b^1 \widetilde{m}_b^2 \widetilde{m}_b^3
      \sum_{l=1}^3 \theta_b^l
      \left( \xi \ln U_l + \zeta \ln T_l \right) \right] \nonumber \\ 
 && - \frac{1}{4} \sum_{i\neq j\neq k=1}^3 n_a^i \widetilde{m}_a^j \widetilde{m}_a^k
   \left[ \sum_b N_b \widetilde{m}_b^i n_b^j n_b^k \sum_{l=1}^3 \theta_b^l
   \left( \xi \ln U_l + \zeta \ln T_l \right) \right] \nonumber \\
 && - \frac{1}{8} \sum_{i;k,l} n_a^i \epsilon_{a,kl}^i \sum_b N_b \widetilde{m}_b^i
    \left( \epsilon_{b,kl}^i - \eta_{\Omega R} \eta_{\Omega R i} \epsilon_{b,R(k)R(l)}^i \right)
    \sum_j \theta_b^j ( \xi \ln U_j + \zeta \ln T_j ) \nonumber \\
 && + \frac{1}{8} \sum_{i;k,l} \widetilde{m}_a^i \epsilon_{a,kl}^i \sum_b N_b n_b^i
    \left( \epsilon_{b,kl}^i + \eta_{\Omega R} \eta_{\Omega R i} \epsilon_{b,R(k)R(l)}^i \right)
    \sum_j \theta_b^j ( \xi \ln U_j + \zeta \ln T_j )\, . \nonumber \\
 \label{univkaehlermet}
\end{eqnarray}
The prime on the first sum indicates that it only runs over branes $b$ that intersect
brane $a$ at generic angles on all three tori.

The first two terms on the RHS of \eqref{univkaehlermet} are cancelled by the correction
arising from the tree-level gauge kinetic function
on the RHS of \eqref{gceffFT} upon redefining the dilaton and complex structure moduli
as follows:
\begin{eqnarray}
 S &\rightarrow& S - \frac{1}{32\pi^2} \sum_b N_b \widetilde{m}_b^1 \widetilde{m}_b^2 \widetilde{m}_b^3
                   \left( \sum_l \theta_b^l ( \xi \ln U_l + \zeta \ln T_l ) \right) \\
 U_i &\rightarrow& U_i + \frac{1}{32\pi^2} \sum_b N_b \widetilde{m}_b^i
             n_b^j n_b^k
             \left( \sum_l \theta_b^l ( \xi \ln U_l + \zeta \ln T_l ) \right)
\end{eqnarray}
In order to interpret the last two terms in \eqref{univkaehlermet} one notices that
the tree-level gauge kinetic function, in addition to the dependence on untwisted moduli
given in \eqref{treelevelgkf}, depends also on the twisted moduli. An anomaly analysis,
sketched in the appendix, suggests that the full tree-level gauge kinetic function is
\begin{eqnarray}
 f_{a, tree} &=& S^c \prod_{i=1}^3 n_a^i - \sum_{i\neq j\neq k=1}^3 U_i^c\, n_a^i \widetilde{m}_a^j \widetilde{m}_a^k
 + \sum_{i=1}^3 \sum_{k,l=1}^4 n_a^i \bigl ( \epsilon^i_{a,kl} 
\nonumber \\ &&
 - \eta_{\Omega R} \eta_{\Omega R i} \epsilon^i_{a,R(k)R(l)} \bigr) W^c_{ikl}
 + \widetilde{m}_a^i \bigl( \epsilon^i_{a,kl} + \eta_{\Omega R} \eta_{\Omega R i} \epsilon^i_{a,R(k)R(l)} \bigr) \widetilde{W}^c_{ikl} ,
 \label{gkftree}
\end{eqnarray}
where $W^c_{ikl}$ and $\widetilde{W}^c_{ikl}$, $i\in\{1,2,3\}$, $k,l\in\{1,2,3,4\}$ are twisted sector fields. There
are $h_{21}$ hypermultiplets coming from the closed string sector in the spectrum of type IIA string theory on a Calabi-Yau space. In
the present case, $h_{21}=h_{21}^{untwisted}+h_{21}^{twisted}$ acquires contributions from untwisted
($h_{21}^{untwisted}=3$) and twisted ($h_{21}^{twisted}=48$) sectors. $W^c_{ikl}$ and $\widetilde{W}^c_{ikl}$
are the $2 h_{21}^{twisted}=96$ complex scalars arising from the sixteen fixed points, labelled by $kl$, in each of
the three twisted sectors, labelled by $i$. The real parts of $W^c_{ikl}$ and $\widetilde{W}^c_{ikl}$ are NSNS-sector fields
and the axions are RR-fields.

The twisted moduli can
also lead to sigma model anomalies in the low energy supergravity theory and can therefore mix
with the dilaton and the other complex structure moduli. One is thus lead to conclude
that the twisted moduli are shifted by
\begin{eqnarray}
 \delta^{(1)} W_{ikl} = 
 - \frac{1}{64\pi^2} \sum_b N_b \widetilde{m}_b^i \epsilon^i_{b,kl}
   \sum_j \theta_b^j (\xi \ln U_j + \zeta \ln T_j)
   \label{shift1}
\end{eqnarray}
and
\begin{eqnarray}
 \delta^{(1)} \widetilde{W}_{ikl} =
 \frac{1}{64\pi^2} \sum_b N_b n_b^i \epsilon^i_{b,kl}
   \sum_j \theta_b^j (\xi \ln U_j + \zeta \ln T_j)
   \label{shift2} ,
\end{eqnarray}
respectively, such that the last two terms in \eqref{univkaehlermet} cancel.

One comment on \eqref{kaehlermetfac2} is in order. From the worldsheet conformal field theory
point of view it is clear that the K\"ahler metrics for the chiral matter arising at a brane
intersection should be the same on the orbifolds with $h_{21}=51$ and $h_{21}=3$. At first
sight it might seem that \eqref{kaehlermetfac2} differs from the corresponding expression
for the other orbifold \cite{Akerblom:2007uc} in that the latter has
$\sign (\prod_i I_{ab}^i)$ in the exponent rather than $\sign (\Upsilon_{ab})$ as in
\eqref{kaehlermetfac2}. There is, however, a physical argument that shows that
these signs must be equal. The orbifold projection removes some
of the string states, but it cannot change their spacetime chirality. As the aforementioned
signs determine the spacetime chirality, they must be equal.

There is one term \eqref{annuuni4} in the gauge threshold corrections
that has so far been neglected.
Upon summing over all annuli with one boundary on brane stack $a$ and
using the tadpole cancellation condition it 
can be cast into
\begin{eqnarray}
 && \sum_i \sum_{k,l} \left( \sum_{b\neq a} N_b I_{ab}^i \epsilon_{a,kl}^i \epsilon_{b,kl}^i (\theta_a^i-\theta_b^i) +
 \sum_{b} N_b I_{ab'}^i \epsilon_{a,kl}^i \epsilon_{b',kl}^i (\theta_a^i-\theta_{b'}^i) \right) \nonumber \\
 &=& \sum_i \sum_{k,l} n_a \epsilon_{a,kl}^i \sum_b N_b \widetilde{m}_b^i \theta_b^i ( -\epsilon_{b,kl}^i +
     \eta_{\Omega R} \eta_{\Omega R i} \epsilon_{b,R(k)R(l)}^i ) \nonumber \\ &&
     + \sum_i \sum_{k,l} \widetilde{m}_a^i \epsilon_{a,kl}^i \sum_b N_b n_b^i \theta_b^i ( \epsilon_{b,kl}^i +
     \eta_{\Omega R} \eta_{\Omega R i} \epsilon_{b,R(k)R(l)}^i ) ,
     \label{universalRD}
\end{eqnarray}
where it was used that
$I_{ab'}^i=-(n_a^i \widetilde{m}_b^i + \widetilde{m}_a^i n_b^i)$,
$\epsilon_{b',kl}^i = -\eta_{\Omega R} \eta_{\Omega R i} \epsilon_{b,R(k)R(l)}^i$ and
$\theta_{b'}^i=-\theta_b^i$. This term resembles the last two terms in \eqref{univkaehlermet}.
One would therefore like to conjecture that this
term is also cancelled by a redefinition of the twisted moduli.
In particular, the shifts would have to be
\begin{eqnarray}
 \delta^{(2)} W_{ikl} &=& 
 - \frac{1}{32\pi^2} \sum_b N_b \widetilde{m}_b^i \epsilon^i_{b,kl}
   \theta_b^i \ln [2] \nonumber \\ 
 \delta^{(2)} \widetilde{W}_{ikl} &=&
 \frac{1}{32\pi^2} \sum_b N_b n_b^i \epsilon^i_{b,kl}
   \theta_b^j \ln [2]
   \label{shiftp}
\end{eqnarray}
Taking into account both the contributions \eqref{shift1}/\eqref{shift2}
and \eqref{shiftp} the real parts of the twisted moduli
acquire the redefinition \eqref{twccmredef}.
Note here that 
\begin{eqnarray}
\sum_{k,l} \epsilon^i_{a,kl}
( \epsilon^i_{b,kl} \pm \eta_{\Omega R} \eta_{\Omega R i} \epsilon^i_{b,R(k)R(l)} ) = 
\sum_{k,l} \epsilon^i_{b,kl}
( \epsilon^i_{a,kl} \pm \eta_{\Omega R} \eta_{\Omega R i} \epsilon^i_{a,R(k)R(l)} ). 
\end{eqnarray}

\section{Summary of  results}
Eventually, even for the danger of repeating ourselves, let us summarise the
main results of this paper. 
The gauge threshold corrections for intersecting D6-brane models
on the $\mathbb{Z}_2\times\mathbb{Z}'_2$ orbifold with $h_{21}=51$,
which allows for rigid three-cycles,
were computed. It was shown that
the results fulfil the non-trivial condition that, in a supersymmetric theory, the gauge kinetic function must
be a holomorphic function of the chiral superfields. Let us emphasise that this is important
because it shows that the D-instanton generated couplings can be incorporated in a holomorphic
superpotential\cite{Akerblom:2007uc}.
For it to be true,
a mixing between the complex structure moduli of the twisted
and untwisted sectors must take place at one loop. In particular,
the twisted complex structure moduli are redefined as 
\begin{eqnarray}
 W_{ikl} \rightarrow W_{ikl} - \frac{1}{64\pi^2} \sum_b N_b \widetilde{m}_b^i
         \epsilon^i_{b,kl}
         \sum_j \theta^j_b \left( \xi \ln U_j + \zeta \ln T_j + \ln [4]\,  \delta_{ij} \right)
        \nonumber \\
 \widetilde{W}_{ikl} \rightarrow \widetilde{W}_{ikl} + \frac{1}{64\pi^2} \sum_b N_b n_b^i
         \epsilon^i_{b,kl}
         \sum_j \theta^j_b \left( \xi \ln U_j + \zeta \ln T_j + \ln [4] \, \delta_{ij} \right).
 \label{twccmredef}
\end{eqnarray}
In contrast to the $\mathbb{Z}_2\times\mathbb{Z}_2$ orbifold with $h_{21}=3$, the gauge kinetic function does in the present
case receive one-loop corrections from sectors preserving $\mathcal{N}=1$ supersymmetry. Upon summing
over all contributions, one finds that the one-loop correction to the gauge kinetic function for the
gauge theory on brane stack $a$ is
\begin{eqnarray}
 f_a^{1-loop} &=& \frac{1}{4\pi} \sum_{i=1}^3 \ln\left[ \eta(i T_i^c)\right]
     -\frac{1}{4\pi^2} \sum_{b\ \in\ case\ 2} \frac{N_b}{\sigma_{ab}} 
     \sum_{i=1}^3 \sigma^i_{ab} \ln\left[ \eta(iT^c_i)\right]  \nonumber \\ &&
     -\frac{1}{8\pi^2} \sum_{b\ \in\ case\ 3} \frac{N_b}{\sigma_{ab}}
                   \sum_{i=1}^3 \sigma^i_{ab} \ln\left[
      \frac{\vartheta
      \genfrac[]{0pt}{}{1/2(1-|\delta_a^i-\delta_b^i|)}{1/2(1-|\lambda_a^i-\lambda_b^i|)}
      (0,iT^c_i)}{\eta(iT^c_i)}\right]\, .
\end{eqnarray}
The K\"ahler metric for the vector-like bifundamental matter arising from strings
stretched between two stacks of branes that are coincident but differ in their twisted charges
was, using holomorphy arguments, determined to be
\begin{eqnarray}
 K_f^{V} = (SU_1U_2U_3)^{-\frac{1}{4}} (T_1T_2T_3)^{-\frac{1}{2}} 
        \left( \prod_{i=1}^3 (V^i_a)^{\sigma_{ab}^i} \right)^\frac{1}{\sigma_{ab}} .
\end{eqnarray}
Equivalently one can determine the K\"ahler metric for
the chiral bifundamental matter arising at the intersection of two stacks of
branes to be
\begin{eqnarray}
 K_{f,ab}^{C} &=& K_{f,ab}^{C(1)} K_{f,ab}^{C(2)} \nonumber \\
&=& S^{-\frac{1}{4}}
\prod_{i=1}^3 U_i^{-1/4-\xi \sign(\Upsilon_{ab})\theta_{ab}^i}
               T_i^{-1/2-\zeta \sign(\Upsilon_{ab})\theta_{ab}^i}
 \times \nonumber \\ &&
  \left[ \prod_{i=1}^3 \left( \frac{\Gamma(1-|\theta_{ab}^i|)}
              {\Gamma(|\theta_{ab}^i|)}\right)^{\sign(\theta_{ab}^i)}
 \right]^{-1/[2\sum_j \sign(\theta_{ab}^j)]}  ,
\end{eqnarray}
where $\xi$ and $\zeta$ are undetermined constants. It was argued \cite{Akerblom:2007uc} that they should be $\xi=0$
and $\zeta=\pm1/2$. These values do however not follow from the calculations performed in this paper.

As already discussed in the introduction, the results of this paper are important
for the study of E2-instantons on the background considered, as the one-loop amplitudes computed here
are equal to one-loop amplitudes in the instanton background. They are therefore
important ingredients  when, e.g., neutrino Majorana masses \cite{Cvetic:2007ku,Cvetic:2007qj} or moduli stabilisation
is studied \cite{Camara:2007dy}.
\vskip 1cm
 {\noindent  {\Large \bf Acknowledgements}}
 \vskip 0.5cm
The authors would like to thank Dieter L{\"u}st, Sebastian Moster, 
Stephan Stieberger and especially Nikolas Akerblom and Erik Plauschinn for valuable discussions.
This work is supported in part by the European Community's Human
Potential Programme under contract MRTN-CT-2004-005104
`Constituents, fundamental forces and symmetries of the universe'.

\begin{appendix}
\section{Anomaly analysis}
Models on the background considered in this paper and described in the main text do
generically have an anomalous spectrum. The anomalies are however cancelled by a
generalised Green-Schwarz mechanism \cite{Aldazabal:1998mr}.

In the following the $U(1)_a - SU(N_b)^2$ anomalies will be considered and it will
be assumed that brane stacks $a$ and $b$ are an example of what was called case 4
in the main text.
\paragraph{{\large Oriented case} \\[8pt]}
It is convenient to start with the case in which there is no orientifold plane.
The anomaly coefficient arising from the chiral multiplets can be computed
to be \cite{Blumenhagen:2005mu}
\begin{eqnarray}
 - N_a \Upsilon_{ab}
 &=& \frac{N_a}{4} \Big[ n_b^1 n_b^2 n_b^3
   \widetilde{m}_a^1 \widetilde{m}_a^2 \widetilde{m}_a^3
   + \sum_{i\neq j\neq k=1}^3 n_b^i \widetilde{m}_b^j \widetilde{m}_b^k 
   \widetilde{m}_a^i n_a^j n_a^k
   \nonumber \\ &&
   - \sum_{i\neq j\neq k=1}^3 \widetilde{m}_b^i n_b^j n_b^k
   n_a^i \widetilde{m}_a^j \widetilde{m}_a^k 
   - \widetilde{m}_b^1 \widetilde{m}_b^2 \widetilde{m}_b^3
   n_a^1 n_a^2 n_a^3
   \nonumber \\ &&
   - \sum_i \sum_{k,l} \widetilde{m}_b^i \epsilon^i_{b,kl} n_a^i
   \epsilon^i_{a,kl}
   + \sum_i \sum_{k,l} n_b^i \epsilon^i_{b,kl} \widetilde{m}_a^i
   \epsilon^i_{a,kl}
   \Big] .
   \label{anomalies}
\end{eqnarray}
The following terms arise from the Chern-Simons actions for the brane stacks $a$ and $b$
\begin{eqnarray}
 S_b^{CS} &\supset&
 \int tr(F_b \wedge F_b) \Bigg[
      n_b^1 n_b^2 n_b^3\,\, A^{(0)}_0
      + \sum_{i\neq j\neq k=1}^3 n_b^i \widetilde{m}_b^j \widetilde{m}_b^k\,\, A^{(0)}_i 
      \nonumber \\ &&
      + \sum_{i\neq j\neq k=1}^3 \widetilde{m}_b^i n_b^j n_b^k\,\, \widetilde{A}^{(0)}_i 
      + \widetilde{m}_b^1 \widetilde{m}_b^2 \widetilde{m}_b^3\,\, \widetilde{A}^{(0)}_0
 \Bigg]
 \\
 S_a^{CS} &\supset&
 N_a \int F_a \wedge \Bigg[
      - n_a^1 n_a^2 n_a^3 \,\, \widetilde{A}^{(2)}_0
      - \sum_{i\neq j \neq k=1}^3 n_a^i \widetilde{m}_a^j \widetilde{m}_a^k\,\, \widetilde{A}^{(2)}_i
      \nonumber \\ &&
      + \sum_{i\neq j\neq k=1}^3 \widetilde{m}_a^i n_a^j n_a^k\,\, A^{(2)}_i
      + \widetilde{m}_a^1 \widetilde{m}_a^2 \widetilde{m}_a^3 \,\,  A^{(2)}_0
 \Bigg]
\end{eqnarray}
and lead to a cancellation of the anomalies described by the first eight summands in
\eqref{anomalies}. $F_a$ is the $U(1)_a$ field strength and $F_b$ the $SU(N_b)$ field strength.
$A^{(0)}_{0,i}$, $\widetilde{A}^{(0)}_{0,i}$ are axions arising from untwisted RR fields and
$A^{(2)}_{0,i}$, $\widetilde{A}^{(2)}_{0,i}$
their 4d dual two-forms.

The remaining anomalies are cancelled if the following couplings of the twisted RR fields
to the gauge fields arise in the low energy effective action \cite{Ibanez:1998qp,Ibanez:1999pw}:
\begin{eqnarray}
 \hat{S}_b^{CS} &=&
 \int tr(F_b \wedge F_b) \Bigg[
      n_b^i \epsilon^i_{b,kl}\,\, A^{(0)}_{ikl}
      + \widetilde{m}^i_b \epsilon^i_{b,kl}\,\, \widetilde{A}^{(0)}_{ikl}
 \Bigg]
 \\
 \hat{S}_a^{CS} &=&
 N_a \int F_a \wedge \Bigg[
      - n_a^i \epsilon^i_{a,kl}\,\,  \widetilde{A}^{(2)}_{ikl}
      + \widetilde{m}^i_a \epsilon^i_{a,kl}\,\,  A^{(2)}_{ikl}
 \Bigg]
 \label{twcoupling}
\end{eqnarray}
Here, $A_{ikl}^{(0)}$ and $\widetilde{A}_{ikl}^{(0)}$ are axions arising from the twisted
RR sectors and $A_{ikl}^{(2)}$, $\widetilde{A}_{ikl}^{(2)}$ their 4d dual two-forms.

\paragraph{{\large Unoriented case} \\[8pt]}
Things are quite similar to the oriented case, but the orientifold images have to be taken into account and
some of the axions are projected out of the spectrum. The anomaly coefficient becomes
\begin{eqnarray}
 \frac{N_a}{2} \left( - \Upsilon_{ab} + \Upsilon_{a'b} \right)
 &=& \frac{N_a}{4} \Big[ n_b^1 n_b^2 n_b^3
 \widetilde{m}_a^1 \widetilde{m}_a^2 \widetilde{m}_a^3
   + \sum_{i\neq j\neq k=1}^3 n_b^i \widetilde{m}_b^j \widetilde{m}_b^k
   \widetilde{m}_a^i n_a^j n_a^k
   \nonumber \\ &&
   + \frac{1}{2} \sum_i \sum_{k,l} \widetilde{m}_b^i \epsilon^i_{b,kl} n_a^i
   ( - \epsilon^i_{a,kl} - \eta_{\Omega R} \eta_{\Omega R i} \epsilon^i_{a,R(k)R(l)} )
   \nonumber \\ &&
   + \frac{1}{2} \sum_i \sum_{k,l} n_b^i \epsilon^i_{b,kl} \widetilde{m}_a^i
   ( \epsilon^i_{a,kl} - \eta_{\Omega R} \eta_{\Omega R i} \epsilon^i_{a,R(k)R(l)} )
   \Big]
   \label{anomaliesU}
\end{eqnarray}
and the Chern-Simons actions yield
\begin{eqnarray}
 S_b^{CS} &\supset&
 \int tr(F_b \wedge F_b) \Bigg[
      n_b^1 n_b^2 n_b^3\,\, A^{(0)}_0
      + \sum_{i\neq j\neq k=1}^3 n_b^i \widetilde{m}_b^j \widetilde{m}_b^k\,\,
 A^{(0)}_i 
 \Bigg]
 \\
 S_a^{CS} &\supset&
 N_a \int F_a \wedge \Bigg[
      \sum_{i\neq j\neq k=1}^3 \widetilde{m}_a^i n_a^j n_a^k\,\,  A^{(2)}_i
      + \widetilde{m}_a^1 \widetilde{m}_a^2 \widetilde{m}_a^3 \,\,  A^{(2)}_0
 \Bigg]
\end{eqnarray}
to cancel the anomalies related to the first four summands in \eqref{anomaliesU}.
Note that $\widetilde{A}_{0,i}$ is projected out, whereas $A_{0,i}$ remains in the
spectrum. Full anomaly cancellation occurs if the couplings
\begin{eqnarray}
 \hat{S}_b^{CS} &=&
 \int tr(F_b \wedge F_b) \Bigg[
      n_b^i ( \epsilon^i_{b,kl} - \eta_{\Omega R} \eta_{\Omega R i} \epsilon^i_{b,R(k)R(l)} ) A^{(0)}_{ikl}
 \nonumber \\ &&
      + \widetilde{m}^i_b ( \epsilon^i_{b,kl} + \eta_{\Omega R} \eta_{\Omega R i} \epsilon^i_{b,R(k)R(l)} ) \widetilde{A}^{(0)}_{ikl}
 \Bigg] \label{twcouplingUp1}
 \\ &=&
 \int tr(F_b \wedge F_b) \Bigg[
      \frac{n_b^i}{2} ( \epsilon^i_{b,kl} - \eta_{\Omega R} \eta_{\Omega R i} \epsilon^i_{b,R(k)R(l)} )
                      ( A^{(0)}_{ikl} - \eta_{\Omega R} \eta_{\Omega R i} A^{(0)}_{iR(k)R(l)} )
 \nonumber \\ &&
      + \frac{\widetilde{m}^i_b}{2} ( \epsilon^i_{b,kl} + \eta_{\Omega R} \eta_{\Omega R i} \epsilon^i_{b,R(k)R(l)} )
                      ( \widetilde{A}^{(0)}_{ikl} + \eta_{\Omega R} \eta_{\Omega R i} \widetilde{A}^{(0)}_{iR(k)R(l)} )
 \Bigg] \label{twcouplingUp2}
\end{eqnarray}
and
\begin{eqnarray}
 \hat{S}_a^{CS} &=&
 N_a \int F_a \wedge \Bigg[
      n_a^i ( - \epsilon^i_{a,kl} - \eta_{\Omega R} \eta_{\Omega R i} \epsilon^i_{a,R(k)R(l)} )\widetilde{A}^{(2)}_{ikl}
 \nonumber \\ &&
      + \widetilde{m}^i_a ( \epsilon^i_{a,kl} - \eta_{\Omega R} \eta_{\Omega R i} \epsilon^i_{a,R(k)R(l)} ) A^{(2)}_{ikl}
 \Bigg] \label{twcouplingU1}
 \\ &=&
 N_a \int F_a \wedge \Bigg[
      \frac{n_a^i}{2} ( - \epsilon^i_{a,kl} - \eta_{\Omega R} \eta_{\Omega R i} \epsilon^i_{a,R(k)R(l)} )
                      ( \widetilde{A}^{(2)}_{ikl} - \eta_{\Omega R} \eta_{\Omega R i} \widetilde{A}^{(2)}_{iR(k)R(l)} )
 \nonumber \\ &&
      + \frac{\widetilde{m}^i_a}{2} ( \epsilon^i_{a,kl} - \eta_{\Omega R} \eta_{\Omega R i} \epsilon^i_{a,R(k)R(l)} )
                      ( A^{(2)}_{ikl}  - \eta_{\Omega R} \eta_{\Omega R i} A^{(2)}_{iR(k)R(l)} )
 \Bigg]
 \label{twcouplingU2}
\end{eqnarray}
are present in the effective action. Note that from \eqref{twcouplingUp2} and \eqref{twcouplingU2}
one can infer which linear combinations of $A_{ikl}$ and $\widetilde{A}_{ikl}$ are projected out.
To be precise, those ones that do not appear in \eqref{twcouplingUp2} and \eqref{twcouplingU2}
are projected out.

\eqref{twcouplingUp1} leads one to conclude
that the combinations $W^c_{ikl}=W_{ikl} + i A_{ikl}^{(0)}$ and
$\widetilde{W}^c_{ikl}=\widetilde{W}_{ikl} + i \widetilde{A}_{ikl}^{(0)}$
(or rather some linear combinations thereof)
are the appropriate complex scalars of the chiral multiplets in
the low energy effective action and that the holomorphic gauge kinetic function
is indeed given by \eqref{gkftree}.

\end{appendix}


\clearpage
\nocite{*}
\bibliography{rev}

\providecommand{\href}[2]{#2}\begingroup\raggedright\begin{thebibliography}{10}

\bibitem{Uranga:2003pz}
A.~M. Uranga, ``Chiral four-dimensional string compactifications with
  intersecting D-branes,'' {\em Class. Quant. Grav.} {\bf 20} (2003)
  S373--S394,
\href{http://www.arXiv.org/abs/hep-th/0301032}{{\tt hep-th/0301032}}.

\bibitem{MarchesanoBuznego:2003hp}
F.~G. Marchesano~Buznego, ``Intersecting D-brane models,''
\href{http://www.arXiv.org/abs/hep-th/0307252}{{\tt hep-th/0307252}}.

\bibitem{Lust:2004ks}
D.~L{\"u}st, ``Intersecting brane worlds: A path to the standard model?,'' {\em
  Class. Quant. Grav.} {\bf 21} (2004) S1399--1424,
\href{http://www.arXiv.org/abs/hep-th/0401156}{{\tt hep-th/0401156}}.

\bibitem{Blumenhagen:2005mu}
R.~Blumenhagen, M.~Cvetic, P.~Langacker, and G.~Shiu, ``Toward realistic
  intersecting D-brane models,'' {\em Ann. Rev. Nucl. Part. Sci.} {\bf 55}
  (2005) 71--139,
\href{http://www.arXiv.org/abs/hep-th/0502005}{{\tt hep-th/0502005}}.

\bibitem{Blumenhagen:2006ci}
R.~Blumenhagen, B.~K{\"o}rs, D.~L{\"u}st, and S.~Stieberger, ``Four-dimensional
  String Compactifications with D-Branes, Orientifolds and Fluxes,'' {\em Phys.
  Rept.} {\bf 445} (2007) 1--193,
\href{http://www.arXiv.org/abs/hep-th/0610327}{{\tt hep-th/0610327}}.

\bibitem{Abel:2006yk}
S.~A. Abel and M.~D. Goodsell, ``Realistic Yukawa couplings through instantons
  in intersecting brane worlds,''
\href{http://www.arXiv.org/abs/hep-th/0612110}{{\tt hep-th/0612110}}.

\bibitem{Akerblom:2006hx}
N.~Akerblom, R.~Blumenhagen, D.~L{\"u}st, E.~Plauschinn, and
  M.~Schmidt-Sommerfeld, ``Non-perturbative SQCD Superpotentials from String
  Instantons,'' {\em JHEP} {\bf 04} (2007) 076,
\href{http://www.arXiv.org/abs/hep-th/0612132}{{\tt hep-th/0612132}}.

\bibitem{Akerblom:2007uc}
N.~Akerblom, R.~Blumenhagen, D.~L{\"u}st, and M.~Schmidt-Sommerfeld,
  ``Instantons and Holomorphic Couplings in Intersecting D- brane Models,''
  {\em JHEP} {\bf 08} (2007) 044,
\href{http://www.arXiv.org/abs/arXiv:0705.2366 [hep-th]}{{\tt arXiv:0705.2366
  [hep-th]}}.

\bibitem{Billo:2007sw}
M.~Billo {\em et al.}, ``Instantons in N=2 magnetized D-brane worlds,''
\href{http://www.arXiv.org/abs/arXiv:0708.3806 [hep-th]}{{\tt arXiv:0708.3806
  [hep-th]}}.

\bibitem{Billo:2007py}
M.~Billo {\em et al.}, ``Instanton effects in N=1 brane models and the Kahler
  metric of twisted matter,''
\href{http://www.arXiv.org/abs/arXiv:0709.0245 [hep-th]}{{\tt arXiv:0709.0245
  [hep-th]}}.

\bibitem{Lust:2003ky}
D.~L{\"u}st and S.~Stieberger, ``Gauge threshold corrections in intersecting
  brane world models,'' {\em Fortsch. Phys.} {\bf 55} (2007) 427--465,
\href{http://www.arXiv.org/abs/hep-th/0302221}{{\tt hep-th/0302221}}.

\bibitem{Akerblom:2007np}
N.~Akerblom, R.~Blumenhagen, D.~L{\"u}st, and M.~Schmidt-Sommerfeld,
  ``Thresholds for intersecting D-branes revisited,'' {\em Phys. Lett.} {\bf
  B652} (2007) 53--59,
\href{http://www.arXiv.org/abs/arXiv:0705.2150 [hep-th]}{{\tt arXiv:0705.2150
  [hep-th]}}.

\bibitem{Dudas:2005jx}
E.~Dudas and C.~Timirgaziu, ``Internal magnetic fields and supersymmetry in
  orientifolds,'' {\em Nucl. Phys.} {\bf B716} (2005) 65--87,
\href{http://www.arXiv.org/abs/hep-th/0502085}{{\tt hep-th/0502085}}.

\bibitem{Antoniadis:1999ux}
I.~Antoniadis, G.~D'Appollonio, E.~Dudas, and A.~Sagnotti, ``Open descendants
  of Z(2) x Z(2) freely-acting orbifolds,'' {\em Nucl. Phys.} {\bf B565} (2000)
  123--156,
\href{http://www.arXiv.org/abs/hep-th/9907184}{{\tt hep-th/9907184}}.

\bibitem{Blumenhagen:2006ab}
R.~Blumenhagen and E.~Plauschinn, ``Intersecting D-branes on shift Z(2) x Z(2)
  orientifolds,'' {\em JHEP} {\bf 08} (2006) 031,
\href{http://www.arXiv.org/abs/hep-th/0604033}{{\tt hep-th/0604033}}.

\bibitem{Blumenhagen:2005tn}
R.~Blumenhagen, M.~Cvetic, F.~Marchesano, and G.~Shiu, ``Chiral D-brane models
  with frozen open string moduli,'' {\em JHEP} {\bf 03} (2005) 050,
\href{http://www.arXiv.org/abs/hep-th/0502095}{{\tt hep-th/0502095}}.

\bibitem{Blumenhagen:2006xt}
R.~Blumenhagen, M.~Cvetic, and T.~Weigand, ``Spacetime instanton corrections in
  4D string vacua - the seesaw mechanism for D-brane models,'' {\em Nucl.
  Phys.} {\bf B771} (2007) 113--142,
\href{http://www.arXiv.org/abs/hep-th/0609191}{{\tt hep-th/0609191}}.

\bibitem{Cvetic:2007ku}
M.~Cvetic, R.~Richter, and T.~Weigand, ``Computation of D-brane instanton
  induced superpotential couplings - Majorana masses from string theory,'' {\em
  Phys. Rev.} {\bf D76} (2007) 086002,
\href{http://www.arXiv.org/abs/hep-th/0703028}{{\tt hep-th/0703028}}.

\bibitem{Derendinger:1991kr}
J.-P. Derendinger, S.~Ferrara, C.~Kounnas, and F.~Zwirner, ``All loop gauge
  couplings from anomaly cancellation in string effective theories,'' {\em
  Phys. Lett.} {\bf B271} (1991)
307--313.

\bibitem{Derendinger:1991hq}
J.~P. Derendinger, S.~Ferrara, C.~Kounnas, and F.~Zwirner, ``On loop
  corrections to string effective field theories: Field dependent gauge
  couplings and sigma model anomalies,'' {\em Nucl. Phys.} {\bf B372} (1992)
145--188.

\bibitem{Kaplunovsky:1995jw}
V.~Kaplunovsky and J.~Louis, ``On Gauge couplings in string theory,'' {\em
  Nucl. Phys.} {\bf B444} (1995) 191--244,
\href{http://www.arXiv.org/abs/hep-th/9502077}{{\tt hep-th/9502077}}.

\bibitem{Cvetic:2007qj}
M.~Cvetic and T.~Weigand, ``Hierarchies from D-brane instantons in globally
  defined Calabi-Yau Orientifolds,''
\href{http://www.arXiv.org/abs/arXiv:0711.0209 [hep-th]}{{\tt arXiv:0711.0209
  [hep-th]}}.

\bibitem{Camara:2007dy}
P.~G. Camara, E.~Dudas, T.~Maillard, and G.~Pradisi, ``String instantons,
  fluxes and moduli stabilization,''
\href{http://www.arXiv.org/abs/arXiv:0710.3080 [hep-th]}{{\tt arXiv:0710.3080
  [hep-th]}}.

\bibitem{Lust:2004cx}
D.~L{\"u}st, P.~Mayr, R.~Richter, and S.~Stieberger, ``Scattering of gauge,
  matter, and moduli fields from intersecting branes,'' {\em Nucl. Phys.} {\bf
  B696} (2004) 205--250,
\href{http://www.arXiv.org/abs/hep-th/0404134}{{\tt hep-th/0404134}}.

\bibitem{Gaberdiel:1999ch}
M.~R. Gaberdiel and J.~Stefanski, Bogdan, ``Dirichlet branes on orbifolds,''
  {\em Nucl. Phys.} {\bf B578} (2000) 58--84,
\href{http://www.arXiv.org/abs/hep-th/9910109}{{\tt hep-th/9910109}}.

\bibitem{Gaberdiel:2000jr}
M.~R. Gaberdiel, ``Lectures on non-BPS Dirichlet branes,'' {\em Class. Quant.
  Grav.} {\bf 17} (2000) 3483--3520,
\href{http://www.arXiv.org/abs/hep-th/0005029}{{\tt hep-th/0005029}}.

\bibitem{Stefanski:2000fp}
J.~Stefanski, Bogdan, ``Dirichlet branes on a Calabi-Yau three-fold orbifold,''
  {\em Nucl. Phys.} {\bf B589} (2000) 292--314,
\href{http://www.arXiv.org/abs/hep-th/0005153}{{\tt hep-th/0005153}}.

\bibitem{Gaberdiel:2000fe}
M.~R. Gaberdiel, ``Discrete torsion orbifolds and D-branes,'' {\em JHEP} {\bf
  11} (2000) 026,
\href{http://www.arXiv.org/abs/hep-th/0008230}{{\tt hep-th/0008230}}.

\bibitem{Quiroz:2001xz}
N.~Quiroz and J.~Stefanski, B., ``Dirichlet branes on orientifolds,'' {\em
  Phys. Rev.} {\bf D66} (2002) 026002,
\href{http://www.arXiv.org/abs/hep-th/0110041}{{\tt hep-th/0110041}}.

\bibitem{Craps:2001xw}
B.~Craps and M.~R. Gaberdiel, ``Discrete torsion orbifolds and D-branes. II,''
  {\em JHEP} {\bf 04} (2001) 013,
\href{http://www.arXiv.org/abs/hep-th/0101143}{{\tt hep-th/0101143}}.

\bibitem{Maiden:2006qe}
J.~Maiden, G.~Shiu, and J.~Stefanski, Bogdan, ``D-brane spectrum and K-theory
  constraints of D = 4, N = 1 orientifolds,'' {\em JHEP} {\bf 04} (2006) 052,
\href{http://www.arXiv.org/abs/hep-th/0602038}{{\tt hep-th/0602038}}.

\bibitem{Blumenhagen:1999md}
R.~Blumenhagen, L.~G{\"o}rlich, and B.~K{\"o}rs, ``Supersymmetric orientifolds
  in 6D with D-branes at angles,'' {\em Nucl. Phys.} {\bf B569} (2000)
  209--228,
\href{http://www.arXiv.org/abs/hep-th/9908130}{{\tt hep-th/9908130}}.

\bibitem{Bachas:1996zt}
C.~Bachas and C.~Fabre, ``Threshold Effects in Open-String Theory,'' {\em Nucl.
  Phys.} {\bf B476} (1996) 418--436,
\href{http://www.arXiv.org/abs/hep-th/9605028}{{\tt hep-th/9605028}}.

\bibitem{Antoniadis:1999ge}
I.~Antoniadis, C.~Bachas, and E.~Dudas, ``Gauge couplings in four-dimensional
  type I string orbifolds,'' {\em Nucl. Phys.} {\bf B560} (1999) 93--134,
\href{http://www.arXiv.org/abs/hep-th/9906039}{{\tt hep-th/9906039}}.

\bibitem{Shifman:1986zi}
M.~A. Shifman and A.~I. Vainshtein, ``Solution of the anomaly puzzle in susy
  gauge theories and the Wilson operator expansion,'' {\em Nucl. Phys.} {\bf
  B277} (1986)
456.

\bibitem{Kaplunovsky:1994fg}
V.~Kaplunovsky and J.~Louis, ``Field dependent gauge couplings in locally
  supersymmetric effective quantum field theories,'' {\em Nucl. Phys.} {\bf
  B422} (1994) 57--124,
\href{http://www.arXiv.org/abs/hep-th/9402005}{{\tt hep-th/9402005}}.

\bibitem{Aldazabal:1998mr}
G.~Aldazabal, A.~Font, L.~E. Ibanez, and G.~Violero, ``D = 4, N = 1, type IIB
  orientifolds,'' {\em Nucl. Phys.} {\bf B536} (1998) 29--68,
\href{http://www.arXiv.org/abs/hep-th/9804026}{{\tt hep-th/9804026}}.

\bibitem{Ibanez:1998qp}
L.~E. Ibanez, R.~Rabadan, and A.~M. Uranga, ``Anomalous U(1)'s in type I and
  type IIB D = 4, N = 1 string vacua,'' {\em Nucl. Phys.} {\bf B542} (1999)
  112--138,
\href{http://www.arXiv.org/abs/hep-th/9808139}{{\tt hep-th/9808139}}.

\bibitem{Ibanez:1999pw}
L.~E. Ibanez, R.~Rabadan, and A.~M. Uranga, ``Sigma-model anomalies in compact
  D = 4, N = 1 type IIB orientifolds and Fayet-Iliopoulos terms,'' {\em Nucl.
  Phys.} {\bf B576} (2000) 285--312,
\href{http://www.arXiv.org/abs/hep-th/9905098}{{\tt hep-th/9905098}}.

\end{thebibliography}\endgroup
\bibliographystyle{utphys}
\end{document}